# Assessing Behaviour Coverage in a Multi-Agent System Simulation for Autonomous Vehicle Testing

Manuel Franco-Vivo*

*University of Bristol, UK*

**Abstract**

As autonomous vehicle technology advances, ensuring the safety and reliability of these systems becomes paramount. Consequently, comprehensive testing methodologies are essential to evaluate the performance of autonomous vehicles in diverse and complex real-world scenarios. This study focuses on the behaviour coverage analysis of a multi-agent system simulation designed for autonomous vehicle testing, and provides a systematic approach to measure and assess behaviour coverage within the simulation environment. By defining a set of driving scenarios, and agent interactions, we evaluate the extent to which the simulation encompasses a broad range of behaviours relevant to autonomous driving.

Our findings highlight the importance of behaviour coverage in validating the effectiveness and robustness of autonomous vehicle systems. Through the analysis of behaviour coverage metrics and coverage-based testing, we identify key areas for improvement and optimization in the simulation framework. Thus, a Model Predictive Control (MPC) pedestrian agent is proposed, where its objective function is formulated to encourage *interesting* tests while promoting a more realistic behaviour than other previously studied pedestrian agents. This research contributes to advancing the field of autonomous vehicle testing by providing insights into the comprehensive evaluation of system behaviour in simulated environments. The results offer valuable implications for enhancing the safety, reliability, and performance of autonomous vehicles through rigorous testing methodologies.

## 1. Introduction

Autonomous vehicle (AV) technology is on the rise, they are no longer a utopia. There are still challenges to overcome, but the use of autonomous vehicles is steadily becoming a reality in some cities [26], and it is expected that the use of automated transportation will expand across the world by 2040 [21]. For instance, robotaxi services are already offered to the public in some large cities both of the USA (e.g. San Francisco, Phoenix) and China (e.g. Beijing, Guanghzou, Shanghai) [16].

According to several studies, more than 90% of traffic accidents are due to human error [3, 24], which could be potentially compensated by decreasing the active driving role of the human user as the automation level is increased [15, 29]. In addition to the expected reduction in traffic-related fatalities, AVs can also favour a more sustainable transportation system, contributing to reducing greenhouse gas emissions, as well as increasing air quality, and resource efficiency [1]. Moreover, human-computer interaction (HCI) in AVs can improve the inclusion and accessibility addressing needs of passengers with disabilities, or other accessibility requirements [7]. Furthermore, the advance and use of AV technology will not only change the mobility system, but can also introduce new business models [10, 16].

---

*Corresponding author.
 E-mail address: mfranco.vivo@gmail.com or pq21381@bristol.ac.uk

Strongly interested in the development and improvement of the driverless technology, governments and technology companies of all countries invest huge funds in research oriented both to fully AVs and incremental development [16].

With an expected $300-400 billion in revenue by 2035 [13], the allure of a safer and more convenient transportation experience remains strong. However, the on-road testing of AV proves to be both expensive and dangerous, and that is where simulation-based verification comes in [14], helping to gain confidence in the trustworthiness of autonomous systems.

In this work, we focus on behaviour coverage in this type of simulation-based testing. Behaviour coverage is the measure of how rigorously a system, such as multi-agent systems (MAS) used in autonomous vehicle simulation, explores and includes a variety of behaviours or scenarios during testing. It is essential in order to assess the effectiveness and reliability of autonomous systems. Greater behaviour coverage implies that the system has encountered a broad range of scenarios, comprised of both common and edge cases, throughout testing. This increases confidence in the system's ability to successfully handle an extensive range of real-world scenarios, thus minimizing the likelihood of unexpected behaviours or operational failures.

From this motivation, the main aim of this work is to study the behaviour of different agents that are part of a scenario for Autonomous Vehicle testing. Since behaviour coverage is a measure of how well a system's behaviour has been tested or evaluated across a diverse range of scenarios and situations, it is essential for validating the performance and reliability of complex systems like autonomous vehicles. Different agent behaviour may help avoid or cause an accident.

To address this goal, we firstly propose behavioural coverage models to complement situation coverage models, by developing multi-agent systems to target these coverage models. These MAS have different strategies, types of agents and agent characteristics. Using an agency directed approach to increase realism of scenarios and generate twice as many effective tests as a pseudo-random approach. Afterwards, we build a testing framework for our models, obtain results, and determine the effectiveness of the different MAS. This allows us to indicate how much the system is exposed to a diverse set of scenarios, both common and edge cases, during testing. Thereby, we can evaluate the reported results according to an analysis of coverage, effectiveness of the tests and the efficiency of test generation to target the respective coverage models.

To that end, this work is organised as follows. In this section, we have introduced the importance of gaining confidence in the trustworthiness of robotic and autonomous systems, in particular of AVs. Section 2 presents an overview of the theoretical basics in the verification and validation of multi-agent system simulation testing, supported by the literature. Section 3 introduces the environment, the different agents, the function employed by the MPC agent and the coverage metrics deployed to analyse the MAS. Results are presented and evaluated in Section 4. Finally, the conclusions are given in Section 5 by summarising key findings.

## 2. Background

The potential benefits of AV technology on human lives have attracted the interest of a number of research projects and papers in the last decade, as well as huge investments valued at $60.3 billion in 2025 which will increase sharply with a compound annual growth rate of 22.2% in the 2025-2035 period [6]. On the other hand, such capabilities bring new challenges in the verification and validation (V&V) and safety assessment of AVs, which is key in order to have this technology successfully deployed at scale, ensuring that their performance meets societal expectations. Basically, AVs are expected to be trusted [8, 25].

In this context, it is worthy to remark that verification is referred to the process of gaining confidence in the system correctness in relation to its requirements, this can be achieved by testing, i.e., by showing that there is no observable difference between the expected and actual behaviours of a system and detecting failures based on its requirements [11, 14].

### 2.1. Coverage

Autonomous vehicles systems, a type of System of Systems (SoS), are composed of multiple subsystems that influence their overall behaviour [8]. The complexity and unpredictability of the environments in which AVs operate, emphasizes the necessity of rigorous testing measures to ensure their reliability and safety, ensuring that autonomous vehicles are tested and simulated across a wide range and diversity of real-world scenarios to reduce the risk of unexpected behaviours or failures in operation. An approach used for software testing to effectively cover a substantial number of potential inputs is coverage criteria, which provides a technique for searching the input space, selecting inputs and stopping testing [4]. Thus, coverage-criteria-based testing can be implemented for the V&V and safety assurance of AV [2, 25], such as the followings:

(*i*) **Scenario coverage.** Considering the requirements for AVs, the term scenario is defined in [28] as "the temporal development between several scenes" which may be established by determined parameters [11, 12]. Scene is defined as "the central interface between perception and behaviour planning & control" [28], and so it consists of all static objects: road network, street furniture, environmental conditions, and a snapshot of any dynamic elements (e.g. other vehicle changing lanes) [11, 12].

(*ii*) **Situation coverage.** This coverage criterion considers the external and internal situations of the AVs which can handle from the situation space [2, 25, 26]. Situation is defined as the space configuration before the simulation [26], and as the subjective conditions and determinants for behaviour at a particular point in time [11, 12]. It is certainly worthy of remark the implementation of the criteria-based metrics as evaluation metrics. In particular, the situation coverage metrics which can be used for obtaining the situations already covered and times each situation has been generated [26].

(*iii*) **Requirements coverage.** The system under test (SUT) is evaluated in line with defined criteria of acceptability [2, 25].

### 2.2. Agency and Multi-Agent Systems

Techniques for maximising the coverage have been studied, which can be classified into four classes: (1) Pseudo-random Generation (PRG); (2) Search-Based Software Testing (SBST); (3) Reinforcement Learning (RL); (4) High Throughput Testing (HTT), see in [25]. Where PRG and SBST are the most commonly used methods [14, 25]. In particular, (*i*) PRG is used to generate tests by pseudo-randomly (to guarantee reproducibility with the same random seeds) covering the state space of inputs or event sequences representing abstract tests, and maximise the coverage by embedding intelligent sampling techniques, which can also detect failures. It is worth noting that constraints are usually imposed, and either constrained pseudo-random test generation or model-based simulation are used for achieving faster coverage [5, 14].

(*ii*) SBST has proved to be an appropriate way of generating test data (even in complex problems) and optimising testing procedure, exploring effective tests data that maximizes the software structure coverage metric [17, 20].

This field is relatively recent, and other new techniques and combinations of the existing ones are being proposed in order to maximise coverage for testing [25]. Moreover no single method can adequately cover a whole system, thus different methods must be used to increase the confidence in the reliability of autonomous systems [14]. Thus, a comprehensive coverage data analysis given in [5] proves that the combined use of constrained pseudo-random and model-based techniques leads to achieve full coverage, outperforming existing approaches such as machine-learning based coverage-directed test generation (CDTG) techniques [18].

As mentioned before, within the automating test generation, machine learning (ML) algorithms, such as RL, have been also optimally implemented in coverage directed-stimulus generation techniques [5, 11, 18]. This subfield of ML is devoted to address the problem faced by an agent that must learn a behaviour by trial-and-error interacting with a dynamic environment [19], and other agents [22]. Thus, the idea of independent coexisting agents led to the extension of multiple agents [27] supported by the potential benefit showed in [30] applied to cooperation. Multiple intelligent agents interacting to deal with complex problems is known as a multi-agent system, which incorporates software agents and their environment. Moreover, the agents can be endowed with different levels of agency: from passive agents (e.g. obstacles), through active agents of low complexity (e.g. pedestrians as part of a group), until rational agents capable of reasoning from a cognitive model, their environment perception and rules for strategic planning and communication in the system according to [11]. In a multi-agent system, agents are independent (i.e., autonomous) and are able to change their behaviour according to environment (self-organisation and self-direction) as they reach their goals.

In verification environment, an AV is considered as a *responder* DUV (Design under Verification), which means that it is a DUV reacting to received stimuli from a sophisticated simulation environment, named *testbenches* [14], while maintaining a correct driving behaviour without breaking traffic rules [11]. In general terms, a testbench is the code implemented to drive a stimulus sequence into a SUT (responder DUV) while observing input protocols, to record coverage and check the response of the SUT, providing a closed environment from the perspective of the SUT [11, 14]. Further details about the components of a CDV testbench can be found in [14] and a example of testbench structure is shown in Figure 1. The responder DUV is expected to be intelligently interactive with its environment (with agents such as humans, other systems, ...), i.e., showing agency, which also implies that the reaction for a given stimulus may vary over time and across different responder DUVs. The introduction of agency into the test environment brings up challenges for responder DUV testing because there is now no predetermined reaction in an interaction sequence. However, agency is required so that a test has the ability to adjust its reaction to the DUV's observed behaviour where such interaction is oriented to achieve the verification goal [14].

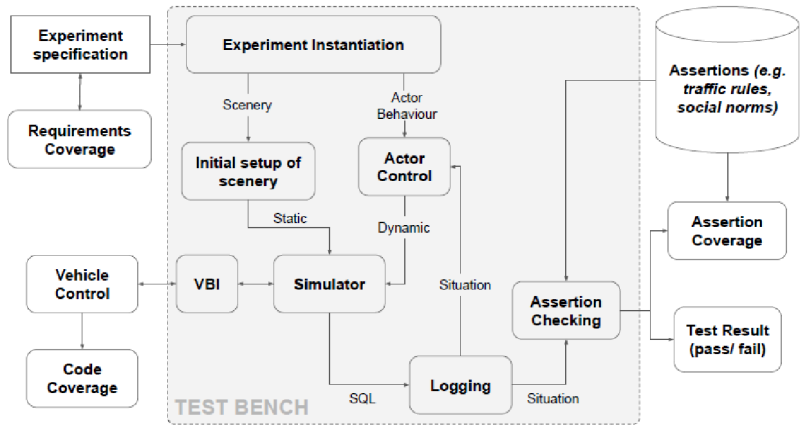

Figure 1: Testbench structure given in [11]

Research focused on studying the benefits of incorporating agency in the verification environment has addressed the issues of verifying the responder DUV by implementing MAS for test generation. In HRI domain, [5] shows that RL can automate Belief-Desire-Intention (BDI) agent models [9] exploration which resulted in more effective code coverage, i.e., higher percentages of automatic code coverage, than the conventional automata-based techniques for model-based test generation [14]. More recently, agency-driven test generation are applied in [11] to provide test agents triggering assertions in simulation-based verification of AVs, such as a decision-making pedestrian crossing a road and testing the reaction time of AV to prevent an impact.

## 3. Methodology

### 3.1. The environment

We adopt a similar test environment to those used in [11] as shown in Figure 2, using CAV-Gym library as a foundation for the environment creation (available at `github/TSL-UOB/CAV-Gym`). We also consider whether a test is interesting or not, that is if there have been pedestrian intrusions in the AV's path in a way that the collision can be avoided by the AV. We can observe this intrusion area in Figure 2 labelled as the preconditioned zone. This is an important measure as these edge cases are essential to achieve faster verification of our AV behaviour. As in [11], we terminate and restart the test when this happens, since this aspect plays a crucial role in evaluating the efficiency of test generation. We do not take into account the test's result (pass or fail) in terms of the AV, as it is not of importance when analysing the behaviour of our test agents.

The environment consists of a straight 99m two-lane road with pavements to each side of the road. The AV travels from left to right, and the pedestrians spawn randomly in the valid spawn locations in the pavement, that is, locations where it is possible for an avoidable collision situation to happen. After each test, the environment is reset and new random initial positions are chosen for each agent, using a fixed seed allows us to control these locations to achieve reproducibility granting us a fair comparison between the different agents.

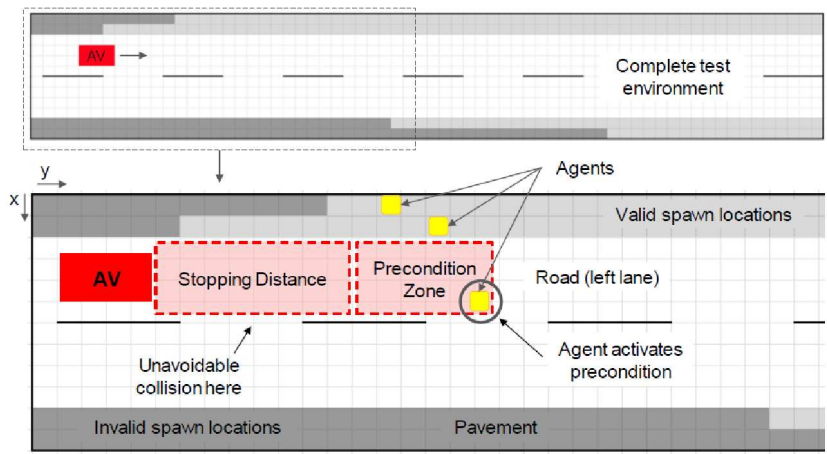

Figure 2: Example of a 2D road in CAV-Gym environment. Source: [11]

*3.2. Test Agents*

The test agents for our different proposed multi-agent systems are the pedestrians that spawn in our environment. We implement the following agent types to analyse how well they target behaviour coverage later on:

- **Random:** A *random agent* makes decisions without taking into account the environment or its current state. It "chooses" actions arbitrarily from available options, often resulting in unpredictable and inefficient outcomes. In our case, as included in CAV-Gym, for a given $\epsilon$ the exploration rate, at each step with probability $\epsilon$ it selects a random action from the action space, and with probability $(1-\epsilon)$ it will select the last action taken (either the default no action taken, or a previously randomly selected action). We use this as a standard against which we can compare the more sophisticated agents.

- **Random Constrained:** They are based on random agents, as they select actions randomly, however these action-selections are done within certain constraints defined by the problem domain. Hence limiting the action space and forcing conditions on the randomness of the actions chosen. In our case, as included in CAV-Gym, at each step for a given $\epsilon$ our agent generates a random number $x$, when $x < \epsilon$ our agent starts crossing the road. We use this agent to introduce structured-randomness into our testing scenarios.

- **Proximity:** They make decisions based on the proximity to other entities in the environment. In our case, as included in CAV-Gym, given a threshold distance $d$, and knowing its own position $(x, y)$ and the car's position $(x_c, y_c)$. If the proximity trigger conditions is satisfied, that is:

$$proximity\_trigger = \begin{cases} \text{True}, & \text{if } \sqrt{((x-x_c)^2 + (y-y_c)^2)} < d \\ \text{False}, & \text{otherwise} \end{cases}$$

then the agent initiates a crossing action. We use this agent since it offers an adaptive approach to decision-making based on spatial relationships, which is specially useful to increase the frequency of interesting tests.

- **Election:** In our case, as included in CAV-Gym, it is based on the proximity agent, when the proximity trigger conditions are satisfied it starts 'voting' for crossing. If voting is initiated, it indicates that the agent is in the process of crossing, of all test agents only one crosses the road. This agent is helpful in introducing agent interaction to the system and increasing realism [11].

- **Q-Learning:** They use reinforcement learning techniques to make optimal decisions in a given environment. They learn by interacting with environment, receiving rewards or penalties for each action, and updating their Q-values accordingly, i.e., a Q-Learning agent trains choosing actions based on predicted maximum rewards [31]. In our case, as included in CAV-Gym, the agent iteratively updates its estimates of the expected rewards for different (state, action) pairs based on observed experiences, gradually learning an optimal policy for decision-making.

- **Model Predictive Control:** An MCP agent is an established control methodology that systematically uses forecasts to compute real-time optimal control decisions. At each time step an optimisation problem is solved over a dynamic environment [23].

In autonomous systems, the ability to make predictions under uncertainty is synonymous with intelligence [23]. Since our autonomous vehicle simulation is a dynamic system, I decided to implement an MPC agent to predict future system behaviour, and optimise a control sequence to minimise a cost function.

*3.3. Formulation*

In this section, we formulate the dynamic optimisation problem in order to minimise the distance between the car and the pedestrian, minimise the agent's time on the road, and minimise the number of changes of direction of the agent in the road. Therefore, we encourage actions that promote edge and interesting tests, while not risking extremely unrealistic pedestrian behaviour.

Now, I define the problem as follows:

> Let $P$ denote the position of the pedestrian, $C$ denote the position of the car, and $t$ denote the time. The objective is to find the optimal trajectory $P(t)$ that minimizes the following cost function:
>
> $$Obj(T) = w_1 \int_0^T \|P(t) - C(t)\|_2 dt + w_2 \int_0^T I(P(t) \in R) dt$$
> $$+ w_3 \sum_{i=1}^{N-1} I\left(\|\theta_{i+1} - \theta_i\|_2 > \epsilon\right)$$
>
> where
>
> - $\|P(t) - C(t)\|_2$ is the Euclidean distance between the pedestrian and the car at time $t$.

- $T$ is the total time horizon.
- $N$ is the number of time steps.
- $\theta_i$ is the direction of the pedestrian at time step $i$.
- $\epsilon$ is a small threshold for considering a change in direction.
- $w_1$, $w_2$, and $w_3$ are weighting factors to balance the importance of each term in the cost function.
- $I()$ is the indicator function.
- $R$ is the road.

This optimisation problem is subject to dynamics constraints and boundary conditions, including constraints on the pedestrian's velocity, acceleration, and permitted changes in direction.

*3.4. Simulation and Metrics*

Each agent behaviour is implemented in Jason [9]. First, we conduct preliminary tests to fine-tune the agent parameters that will be incorporated into our testing framework. Then, during the testing simulation, we record the agents' scores given by a simple scoring system in CAV-Gym (based on realism of the agents behaviour and the success of encountering edge cases), time to test completion and some coverage metrics, which will be explained below, in a log. For each agent type, we build a system of $n$ test agents, $n \in [1,5]$. Each multi-agent system is run 5 times, for 100 tests each run. The coverage metrics used to evaluate behaviour coverage for each MAS are:

- **Scenario coverage metric:** as defined in Section 2.1, a snapshot of all constant and dynamic elements, we evaluate how diverse or how many scenarios are covered by the model.

- **Situation coverage metric:** as defined in Section 2.1, the space configuration before simulation, and the conditions for behaviour at a certain point in time, we implement a metric which obtains how many situations have already been generated by the model.

- **Agent action coverage metric:** this metric helps identify how each agent reacts to the environment at a certain time, giving an insight on how the agents behave based on their surroundings.

## 4. Results and Evaluation

The reported results are evaluated according to the following criteria mentioned in Section 2.1.

*4.1. Scenario coverage results*

After analysing the results for Scenario coverage, we are drawn to the conclusion that, although Scenario and Situation coverage are not equivalent, in our empirical study, Situation coverage analysis encompasses Scenario coverage as well as providing more relevant information about the testing. It produces a more complete set of results, and so Scenario coverage results do not add extra information to our study.

*4.2. Situation coverage results*

From Figure 3, we are able to observe that for smaller number of agents, the higher the decision-making ability of the agent the smaller the Situation coverage ratio. This is because as we will discuss later on, in Figure 5, for smaller numbers of agents a higher agency level implies a greater number of interesting tests. This means that our tests are terminated earlier because they are found interesting, thus for higher agency levels less situations are being generated when the number of agents is small. However, we also find that at higher agent numbers ($n = 5$), where the number of interesting results produced per run becomes closer for each agent type, MPC and Q-Learning agents have higher Situation coverage, showing more realistic behaviour in both common and edge cases, than other random or lower level directed agents.

Random agent has such high Situation coverage ratio because of its random nature, meaning that after running a hundred tests five-times, it is more likely to behave differently (compared to the other agent types) each time, generating a higher total Situation coverage ratio. This shows both a less realistic behaviour (erratic movement which is not normally displayed by real-life pedestrians) and lower success of producing interesting tests. Another feature that should be commented is that Random Constrained agent, which we would expect to have similar Situation coverage to the Random agent, is closer to some of the lower-level directed agent behaviours such as Proximity and Election, this is because a low $\epsilon$ was chosen, encouraging this agent type to cross the road and allowing it to produce a higher number of interesting tests at the expense of a less realistic behaviour.

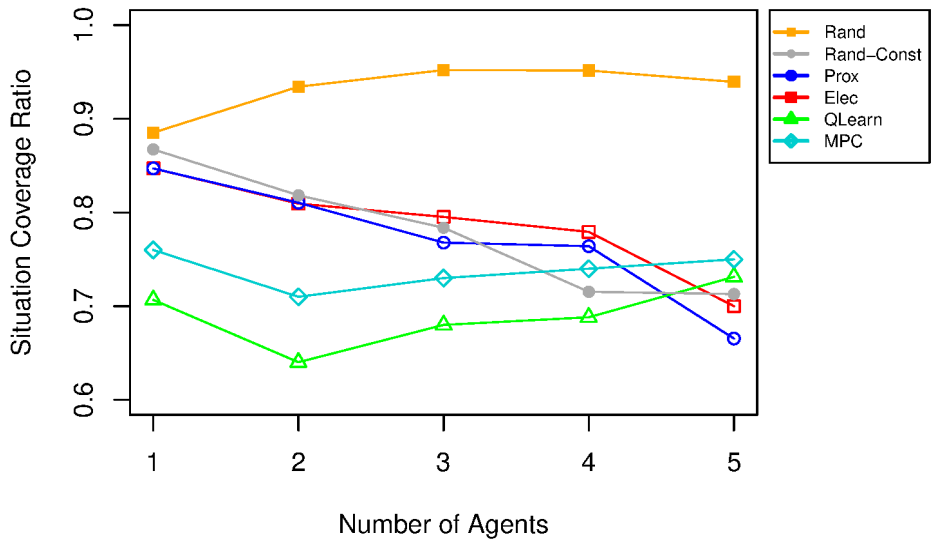

Figure 3: Evolution of the Situation coverage results according to the agent number for each agent type.

*4.3. Agent action coverage metric*

In Figure 4a we observe that the Action Coverage across tests for Random agents is very high compared to the rest, as in the previous coverage metric, this is due to its random nature causing it to exhibit random behaviour across the different iterations of the tests. Focusing on the rest of the types, Figure 4b, we observe that MPC and Q-Learning agents which have a higher level of freedom, i.e. take 'intelligent' decisions before every action, display a higher Action coverage for each test with $n$ agents, $n \in [1,5]$. Random-Constrained has a lower Action coverage, as the small $\epsilon$ used causes the agent to cross the road more frequently, limiting its freedom.

The lower-level directed agents display the lowest Action coverage, their action decision-making, based on responding to proximity based stimuli, restricts the diversity of actions compared to the other agents.

*4.4. Evaluation*

In addition, the evaluation is carried out according to the following three criteria:

- **Accuracy** is the percentage of tests that were interesting from the total amount of tests.

- **Score** is a metric of how natural the agents behaved.

- **CPU time** is a measure of the computational execution cost (of the global computational test)

*4.4.1. Accuracy results*

The percentages of Accuracy results for each $n$ agents, $n \in [1,5]$ are plotted in Figure 5. Here, we have that for small $n$, the accuracy is sorted according to the agent's decision-making capacity, with Q-Learning agents having the most accuracy and Random agents the least. Afterwards, the Proximity agent demonstrates a higher accuracy than both MPC and Election agents, since unlike them it focuses on interesting test generation leaving realistic behaviour aside. Meanwhile random agents' (Random and Random-Constrained) accuracy is purely based on the number of agents due to their arbitrary nature, a larger number of agents implies a higher probability of randomly generating an interesting test.

*4.4.2. Score results*

Figure 6 depicts the averages of the Score results, which represent a measure of how realistic the behaviour of our agent model is. We observe that as the number of agents increases, the difference in the extent to which the agents' behaviour reflects real-world behavior between directed and random agents increases. It shows the importance of using directed agency approach instead of random agency when aiming for realistic behaviour. It can also be seen that directed agents with a higher degree of freedom such as MPC and Q-Learning agents achieve a more natural behaviour than Election and Proximity agents. Specifically, MPC agents, which take into account erratic behaviour, like constant changes in direction or long times in road, inside the cost function it optimises, manifest the most natural behaviour.

*4.4.3. CPU times*

Figure 7 displays the resource consumption needed by each agent system to carry out actions. The resource cost to execute MPC agent actions is around two times the time needed for the second most costly (Q-Learning agents). This is due to the complexity of solving the optimisation problem over a finite time horizon in an environment with nonlinear

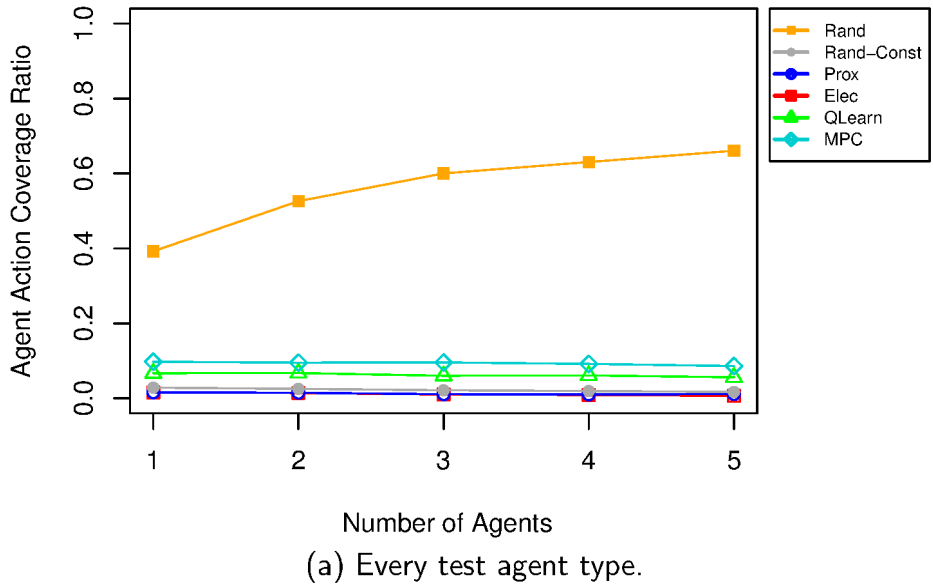

(a) Every test agent type.

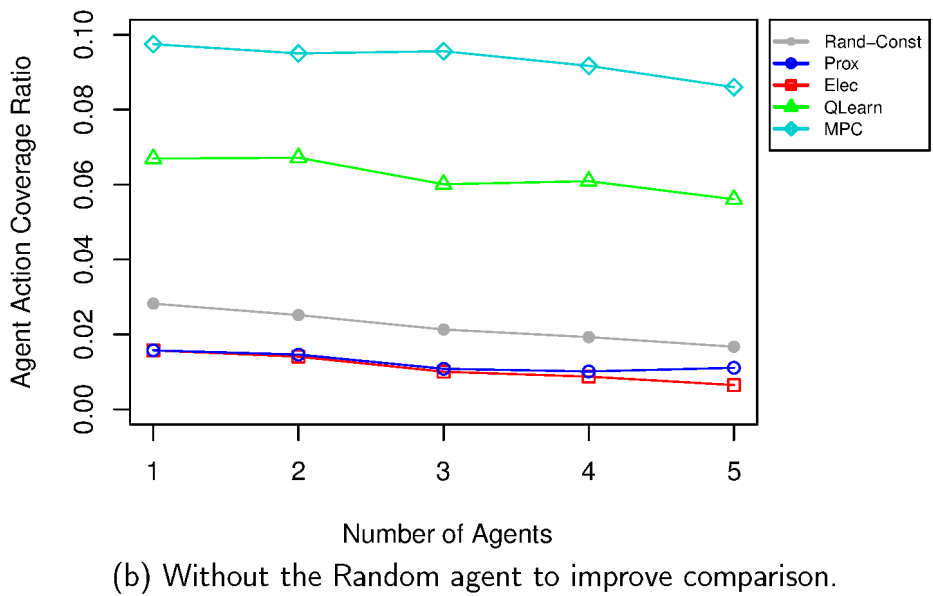

(b) Without the Random agent to improve comparison.

Figure 4: Evolution of the action coverage results according to the agent number.

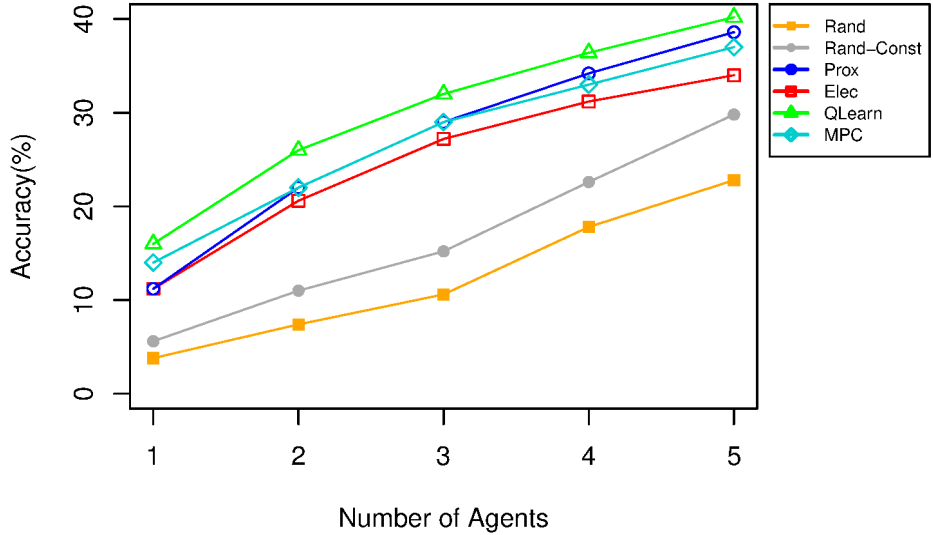

Figure 5: Evolution of the percentage of Accuracy results according to the agent number for each agent type.

dynamics. Basically agents with a higher decision-making power, i.e., higher autonomy level demand more computational resources.

The empirical results for these three metrics have been summarised in Table 1, which corresponds to one agent and five agents.

Overall, we can assess effectiveness and efficiency of our proposed agent-based models from Figure 5 and Table 1, from which we observe that any of the four agency-directed models proposed are effective and efficiency in test generation. Specially, Q-Learning, Proximity, and MPC multi-agent systems prove to have acceptable effectiveness in generating tests. We also evaluate behaviour coverage of our models from Figures 3, 4a and 6, from which we obtain that Q-Learning and specially MPC are the models with a more suitable behaviour coverage, tests generated using these models have greater real-world applicability. To transfer these evaluations to a broader perspective, we consider the computational cost of these multi-agent-based models. We have that for the previously mentioned most suitable models, MPC had a larger computational cost than Q-Learning.

| Agent type | Description | $n = 1$ | | | $n = 5$ | | |
|---|---|---|---|---|---|---|---|
| | | ACC (%) | $S \times 10^4$ | $t_{\text{CPU}}$ (ms) | ACC (%) | $S \times 10^4$ | $t_{\text{CPU}}$ (ms) |
| Random | Select a random action from the action space | 3.8000 | −4.2341 | 0.7177 | 22.8000 | −20.6095 | 8.1546 |
| Constrained random | Select a random action within a restricted action space | 5.6000 | −6.0065 | 0.7101 | 29.8000 | −26.7190 | 7.9186 |
| Proximity | Select an action according to a threshold distance | 11.2000 | −2.9940 | 2.8230 | 38.6000 | −8.2536 | 10.4120 |
| Election | As a proximity agent with proximity-trigger conditions | 11.2000 | −2.9940 | 2.8217 | 34.0000 | −6.2533 | 10.9014 |
| Q-Learning | Use RL to make optimal decisions | 16.0000 | −0.6311 | 4.9954 | 40.2000 | −4.5633 | 13.2895 |
| MPC | Predict system behaviour and optimise a control sequence to minimise a cost | 14.1000 | −0.5741 | 10.3012 | 37.0000 | −3.7102 | 24.5700 |

Table 1: Test agent summary displaying agent type, description, and evaluation metrics for one agent ($n = 1$) and five agents ($n = 5$): accuracy (ACC%), score ($S \times 10^4$), CPU time ($t_{\text{CPU}}$ in ms).

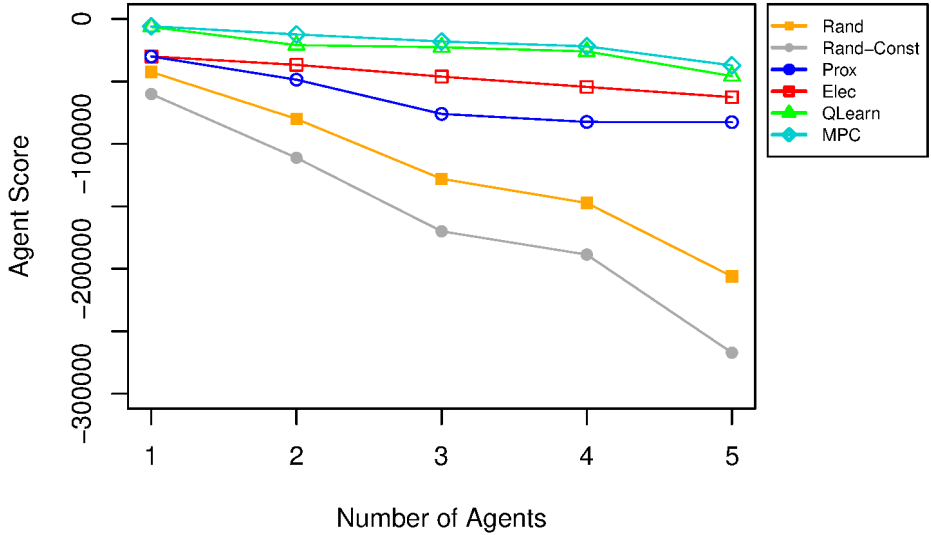

Figure 6: Evolution of the percentage of average Score results according to the agent number for each agent type.

## 5. Conclusion

Multi-agent systems simulation provide a comprehensive and flexible framework for testing and evaluating autonomous vehicle systems in realistic, dynamic and challenging environments, accelerating the development and improving the performance of safe and reliable autonomous vehicles.

In this context, this work was aimed at expanding the study of behaviour of different agents in an Autonomous Vehicle testing scenario. It has been achieved through a behaviour coverage analysis of multi-agent systems simulation designed for autonomous vehicle testing, in which we defined a set of driving scenarios, and agent interactions.

With respect to the objectives proposed in Section 1, we have included general autonomous systems as test agents in the testing framework, increasing realism of scenarios by penalising erratic behaviour. Proposing a dynamic optimisation model for our MPC agent with ability to forecast over a predictive horizon. We have built a testing framework for six different models: Random, Random-Constrained, Proximity, Election, Q-Learning and MPC, including our MPC agent. From which we have obtained results including metrics to assess the behaviour coverage, effectiveness and efficiency and computational cost. We explored and compared the behavioural coverage of these models to assess their effectiveness in simulating real-world pedestrian conduct. From our results:

- We have determined that both Q-Learning and MPC are the models that demonstrate a trade-off between realistic behaviour and effective test generation. While MPC agents display a more authentic behaviour resembling human decision-making, Q-Learning agents offer a notable advantage in terms of computational efficiency, requiring only half the computational resources.

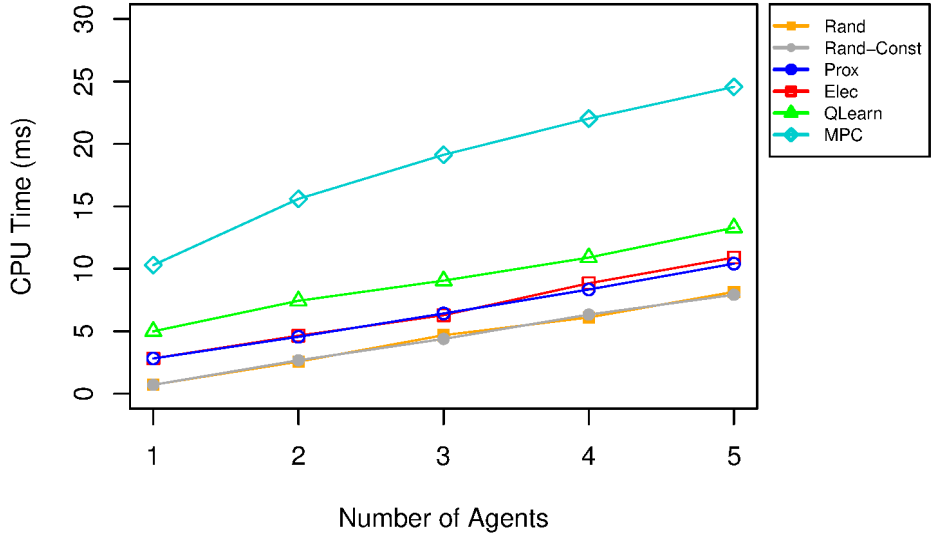

Figure 7: Evolution of the percentage of Average time results according to the agent number for each agent type.

- The choice between Q-Learning and MPC ultimately depends on the specific priorities of the application. If the primary goal is to prioritize realism and human-like behaviour, MPC agents offer a superior solution. On the other hand, if cost-effectiveness and computational efficiency are topmost concerns, opting for Q-Learning agents would be more advantageous.

Finally, further work could be directed towards the implementation of genetic algorithms for MPC optimisation. By using genetic algorithms, the process of searching for the optimal control sequence can be parallelised and distributed across multiple cores. This parallelization enables quicker convergence to solutions that satisfy performance criteria, potentially resulting in reduced computational time compared to conventional optimization techniques.